\documentclass[pre,aps,twocolumn,floatfix,showpacs]{revtex4-2}
\usepackage{graphicx}
\usepackage{dcolumn}
\usepackage{bm}
\usepackage{color}
\usepackage{amssymb}
\usepackage{amsmath}
\usepackage [latin1]{inputenc}
\usepackage{float}
\usepackage{tikz}
\begin{document}

\title{Discriminating protein tags on a dsDNA construct using a Dual Nanopore Device}

\author{Swarnadeep Seth$^1$}
\author{Arthur Rand$^4$}
\author{Walter Reisner$^2$}
\author{William B. Dunbar$^4$}
\author{Robert Sladek$^3$}
\author{Aniket Bhattacharya$^1$}

\altaffiliation[]
{Author to whom the correspondence should be addressed}
{}
\email{AniketBhattacharya@ucf.edu}
\affiliation{$^1$Department of Physics, University of Central Florida, Orlando, Florida 32816-2385, USA}
\affiliation{$^2$Department of Physics, McGill University, 3600 rue university, Montreal, Quebec H3A 2T8, Canada}
\affiliation{$^3$Departments of Medicine \& Human Genetics, McGill University, Montreal, H3A 0G1, Canada}
\affiliation{$^4$Nooma Bio, 250 Natural Bridge Dr, Santa Cruz, CA 95060, USA}

\date{\today}

\begin{abstract}
  We report a novel simulation strategy that enables us to identify key parameters controlling the experimentally measurable characteristics of structural protein tags on dsDNA construct translocating through a double nanopore setup. First, we validate the scheme {\em in silico} by reproducing and explaining the physical origin of the experimental dwell time distributions of the Streptavidin markers on a 48 kbp long dsDNA. These studies reveal the important differences in the characteristics of the protein tags compared to the dynamics of dsDNA segments in between the motifs, immediately providing clues on how to improve the measurement protocols to decipher the unknown genomic lengths accurately. Of particular importance is the {\em in silico} studies on the effect of electric field inside and beyond the pores which we find is critical to discriminate protein tags based on their effective charges and masses revealed 
through a generic power-law dependence of the average dwell time at each pore. The simulation protocols enable to monitor piecewise
dynamics of the individual monomers at a sub-nanometer length scale and provide an explanation of the disparate velocity variation from one tag to the other using the nonequilibrium tension propagation theory, - a key element to decipher genomic lengths accurately. We further justify the model and the chosen simulation parameters by calculating the P\'{e}clet number which is in close agreement with the experiment. Analysis of our simulation results from the CG model 
has the capability to  refine the accuracy of the experimentally obtained genomic lengths and carefully chosen simulation strategies can serve as a powerful tool to discriminate different types of neutral and charged tags 
of different origins on a dsDNA construct in terms of their physical characteristics and can provide insights to increase both the efficiency and accuracy of an experimental dual-nanopore setup.
\end{abstract}
\maketitle
When a biopolymer is driven through a nanopore under an applied electric field, the molecule's passage creates a dynamic modulation of the trans-pore ionic current that can be used to deduce chemical, structural, and conformational properties of the translocating polymer \cite{Review0, Review1,Review2,Review3}. Biopolymer transport through nanopores offers significant prospects for human health.  Pores based on modified transmembrane proteins form the basis of a powerful label free DNA sequencing technology  \cite{Review2}, and there is hope that solid-state pores (ss-pores) based on 2D nanomaterials \cite{farimani1} may eventually sequence with sufficiently high resolution to directly read current fluctuations from one nucleotide at a time passing through the pore.
\par
In addition to their potential thinness, ss-pores are attractive because they can be fabricated with larger dimensions ($>$2.5\,nm diameter) suitable for analyzing not just the translocation of pure ss- and ds- DNA, but DNA with bound molecular features of nanometric size that function as physical tags or have intrinsic biological functionality \cite{albrecht}. A wide-range of features on dsDNA have been detected with ss-pores, including  proteins such as streptavidin labels \cite{chen1, kong1}, anti-DNA antibodies \cite{plesa1}, DNA hairpins \cite{bell1,chen2}, protein nucleic acids \cite{meller2} and aptamers \cite{sze1, kong2}. Such bound molecular features can be readily detected as these features give rise to a secondary blockade riding on top of the underlying DNA blockade in the measured trans-pore ionic current.  The duration (dwell-time) and amplitude of these feature associated blockades contain information concerning the physical characteristics of the tags. If the molecule translocation is linear (i.e. no folds are present), the feature blockades can also provide information regarding the feature's binding position with respect to the molecule's underlying sequence, or the relative distance of the given feature from other features  \cite{chen1, bell1}.  For example, molecular features that bind specifically to repetitive sequence motifs (for example, the recognition sequence of nicking endonucleases), produce a barcode that can then be aligned genome scale \cite{nooma4}. In proposed DNA information storage applications, molecular features can be used to represent the position of `1' bits \cite{chen2}. In functional genomics applications, there is a need to map a wide-range of overlapping transcriptional control mechanisms, for example arising from modified bases \cite{modbaserev} and histone marks \cite{histmaprev}.
\par
\begin{figure}[ht!]
\includegraphics[width=0.48\textwidth]{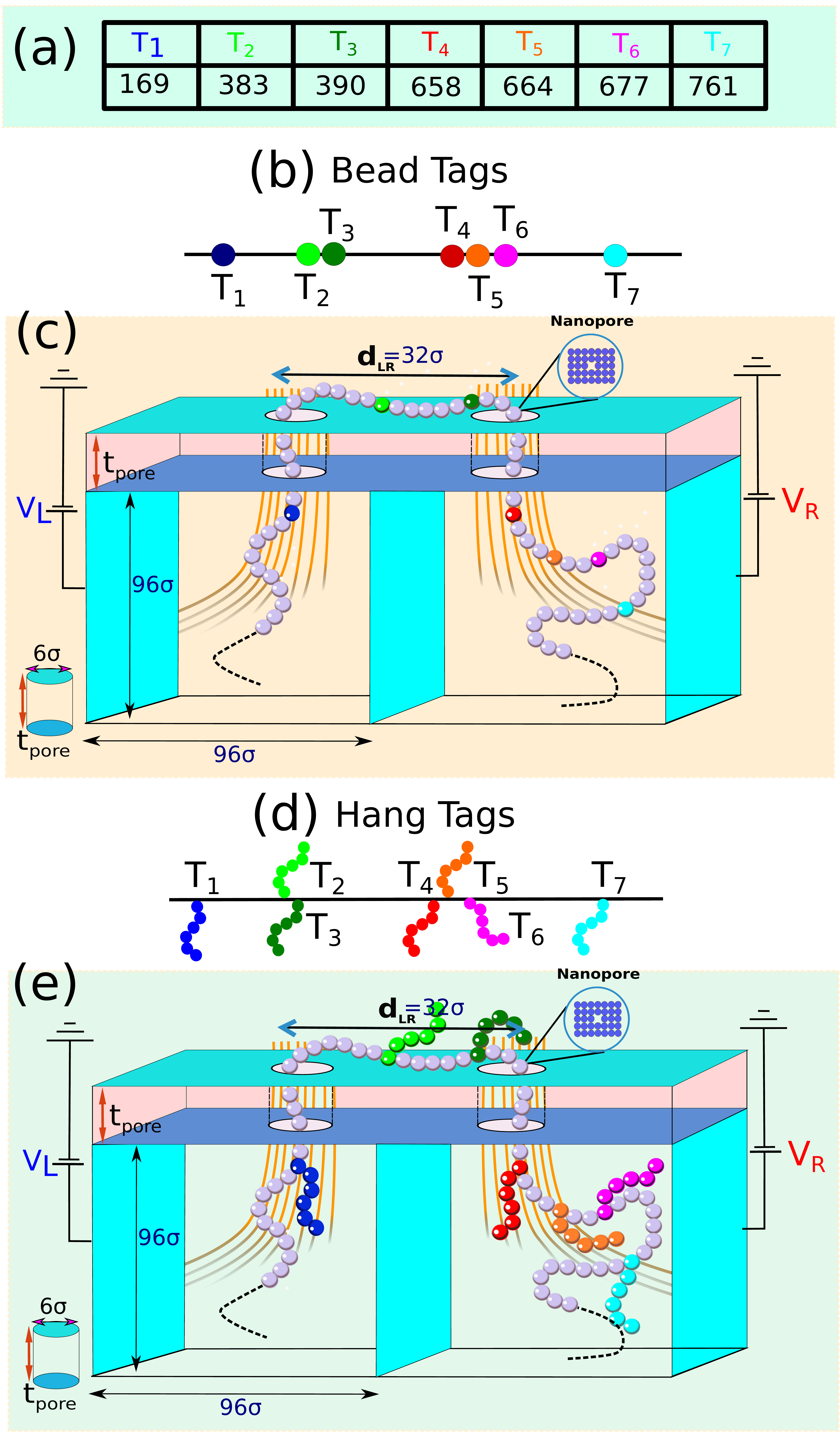}
\caption{\small Schematics of the dual nanopore set up. (a) The simulated DNA construct with locations of the seven tags in a chain of length $L=1024\sigma$, where $\sigma$ is the diameter of each monomer of mass $m$. (b) The tags are modeled in the form of heavier beads of mass $6m$ (c) Simulated dual nanopore system: two nanopores of slightly different diameters $7\sigma$ (left pore)  
and $6\sigma$ (right pore) respectively separated by a distance $d_{LR} = 32\sigma$ are connected to two reservoirs of width $96\sigma$. A spatially extended electric field is applied in the cavity-nanopore system. The voltage at the right pore $V_R$ is kept at a constant value while the voltage at the left pore oscillates $V_L = V_R \pm \Delta V$ after every scan of the DNA which translocates from the left pore and vice-versa. The electric-field profile is calculated using the Finite-element method and the normalized color map shows the field strength (not in scale). (d) The sidechain motifs are modeled with with six monomers. (e) same as in (c) replacing beads by side-chains.}
\label{Model}
\end{figure}

 
A core challenge in nanopore based feature sensing is developing techniques for accurate feature mapping and performing effective discrimination of different feature types.  The double nanopore platform may have potential to outperform devices based on single nanopores in this respect~\cite{Dekker-2016,Aksimentiev-2020,Small1,Small2,Small3}.    If a molecule is simultaneously captured at both pores in a dual pore device, applying opposing biasing to the pores will capture the molecule in a tug-of-war state where the molecule is extended between the pores~\cite{TOW} (Fig.~\ref{Model}).  Such a state suppresses folding giving rise to predominantly linearized translocation traces~\cite{Small2}. In addition, if biasing can be independently adjusted at each pore, the two pore device can achieve controlled slow-down while maintaining high signal-to-noise current sensing~\cite{Small2}. This arises because the threading speed of the molecule between the pores is controlled by the difference in the potential biasing applied to each pore, while the signal is determined by the absolute bias level at each pore. Finally, when coupled to active logic that enables feedback between current measured at the pores and pore biasing, bipolar scanning can be achieved via flipping the differential bias~\cite{Small3}; this enables repetitive scanning of a given genomic region (termed `flossing'),~\cite{Dekker-2016,Aksimentiev-2020,Small1,Small2,Small3} enabling us to deduce the statistical distribution of feature characteristics and improve the measurement's statistical accuracy.  
\par
In this communication, we demonstrate that dual-pore translocation of tagged DNA in a tug-of-war regime has an intrinsic asymmetry: the dwell time of features passing through the entrance pore is on average higher than the dwell time at the exit pore.  Using simulation performed using a coarse-grained (CG) model of the DNA-tag system used in dual-pore flossing experiments, we demonstrate that this aspect of dual-pore translocation is more than just a curiosity, but in fact arises from physically distinct translocation physics specific to the dual-pore platform and may assist in discriminating between different feature classes. In a single pore device, the molecule always translocates in a direction aligned with the local electrophoretic force exerted at the pore.  However, in a dual pore device with opposing biases applied to the pores, translocation can take place at the entrance pore in a direction opposite that of the local electrophoretic force exerted at this pore.  In particular, the local electrophoretic force at the entrance pore acts to slow-down the passage of DNA through the entrance relative to the exit pore.  When charged features are present on the DNA, the effect is to increase the passage time of the features through the entrance pore relative to the exit pore. The simulation results additionally reveal the subchain conformations and the dynamics in between the molecular features hard to access experimentally and bring out the subtleties of the interplay of the electric field inside and in the vicinity of the pore. This includes the evolution and unfolding of the non-equilibrium chain conformations due to reversal of the electric field and the flossing direction; the inertial effect and the effect of the tension propagation (TP) along the chain backbone~\cite{Sakaue_PRE_2007,Ikonen_JCP2012,
Adhikari_JCP_2013,Keyser_NatCommn2017,Keyser_NatPhy2021} and in particular how the propagation dynamics is affected by the presence of the tags that temporarily halt the tension propagation.  The simulation results nicely capture the characteristics of different types of tags, neutral and charged, extended or localized, and display insights and physical understanding of the translocation process. Furthermore, a variation of the simulation parameters beyond those used in experiments
enable us to understand the physical origin of the experimental uncertainties and more efficient design and analysis protocols of future multi-nanopore platforms.
\par
{\bf Simulation \& Model Details:}~Fig.~\ref{Model} shows the schematics of the simulated system. A semiflexible chain of $N=1024$ monomers (beads) of diameter $\sigma$ is used to model a  48500 base pair (bp) long $\lambda$-phage dsDNA construct of contour length $L =N\sigma \approx 16.5\;\mu\rm{m}$ as used in the 
original experiment~\cite{Small3}. This translates to 
$\sigma \approx 47\;{\rm bp} \simeq 16\; {\rm nm}$ for each bead (monomer). In the actual experiment seven tags are placed along the chain whose relative positions are shown in Fig.~\ref{Model}(a). The tags in the experiment consist of monomeric Stretpavidin bound ssDNA of 75 bp~\cite{Small3}. The charge of each bead of the dsDNA is chosen to be unity and the combined monstreptavidin-ssDNA labels are partially charged. In order to understand the entropic and inertial effects we have used two types of tags in the simulation. Fig.~\ref{Model}(b) shows {\em spherical tags} in the form of beads of same diameter $\sigma$ but having a mass $m_{bead}=6m$, where $m$ is the mass of the individual chain monomers. Fig.~\ref{Model}(d) shows {\em side-chain} consisting of 6 monomers, each of mass $m_{\rm sidechain} = m$ and of the same diameter $\sigma$. We have also considered both {\em neutral} and {\em partially charged tags} in order to
understand the effect of the electric field on the charged labels. The dsDNA is co-captured in (Fig.~\ref{Model}(c) and (f)) two closely spaced nanopores drilled on a common membrane and a tug-of-war situation is created by applying voltages $V_L$ and $V_R$ to the left (L) and right (R) reservoirs as shown. In compliance with the experiment we have chosen slightly different pore diameters of $6\sigma$ for the left pore and  $7\sigma$ for the right pore respectively. The distance between the
pores in the actual experiment is $550$ nm which translates to $d_{\rm LR}=32\sigma$ in our simulation. We have also studied the cases for the symmetric L and R pores where $d_L^{pore} = d_R^{pore}= 6\sigma$ and observed that a 10-20\% asymmetry in pore diameters does not make a large qualitative change.
\par
\begin{figure}[ht!]
\includegraphics[width=0.45\textwidth]{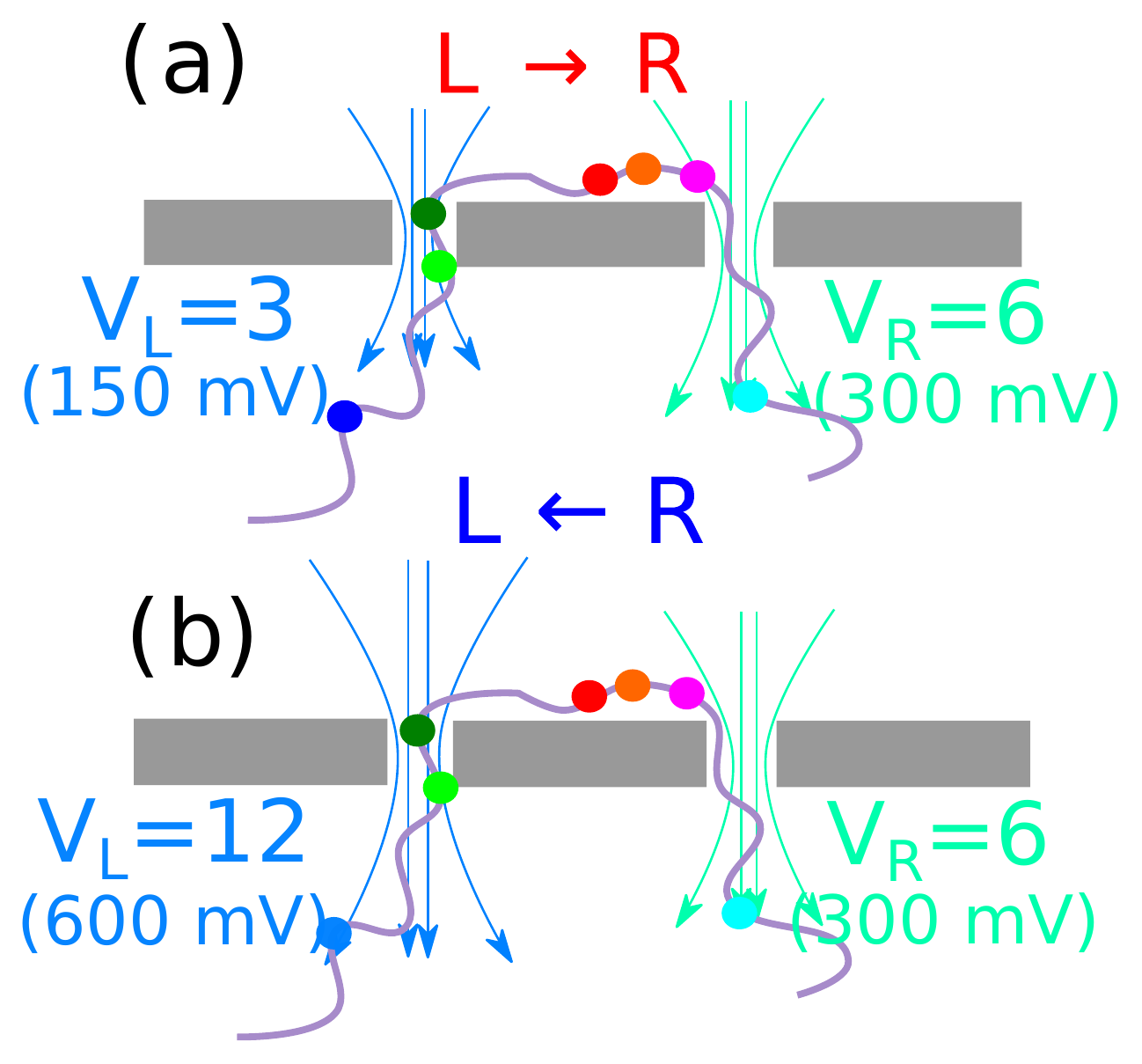}
\caption{\small The schematic of the applied voltage to floss the co-captured dsDNA through the dual nanopore device. The voltage at the right pore  $V_R$ is kept constant while the voltage at the left pore $V_L = V_R + \Delta V_{L\rightarrow R}$ for ${L\rightarrow R}$ translocation and 
$V_L = V_R + \Delta V_{R\rightarrow L}$ for ${R\rightarrow L}$ translocation.   The ratio of the voltages $V_L/V_R=0.5$ and 1.5 for ${L\rightarrow R}$ and ${R\rightarrow L}$ translocations are the same as in the experiment~\cite{Small3}(Fig.~\ref{Dwell}(a)-(b)). $\Delta V_{L\rightarrow R} \ne \Delta V_{R\rightarrow L}$ in general. Other combinations are studied in Fig.~\ref{Dwell}(c)-(d) through Fig.~\ref{Dwell} (k)-(l).  
}
\label{E-Field}
\end{figure}

{\bf Flossing the captured dsDNA and the electric field in and around the pore:}
It is important to understand how the biases are applied in each
pore in order to scan the dsDNA multiple times. In the experimental
protocol ~\cite{Small3} the voltage across the right pore $V_R=150$ mV is kept
constant while the voltage across the left pore $V_L$ is switched from
300 mV to $600 \pm 50$ mV for translocation to occur from the $L \rightarrow R$
and $R \rightarrow L$ respectively. We have used similar ratios for in
the simulation and varied the bias at the left pore only as shown in Fig~\ref{E-Field}.
We have translated 50 mV to one unit of applied bias so that the biases 150 mV, 300 mV, and 600 mV translates to 3, 6, and 12 simulation units (Fig.~\ref{E-Field}).
This choice is later justified to reproduce similar P\'{e}clet number $\simeq 50$ for the simulation as well as the experiment. 
The co-captured dNA is scanned repeatedly by
altering the voltage bias $V_L$ at the left pore only while keeping
the bias at the right pore $V_R$ the same. Thus, $V_L =  V_R + (\Delta
V)_{L\rightarrow R} < V_R$ for the $L \rightarrow R$ translocation. Likewise, $V_L =  V_R - (\Delta
V)_{R\rightarrow L
} > V_R$ for the $R \rightarrow L$ translocation.
\begin{figure}[ht!]
\includegraphics[width=0.46\textwidth]{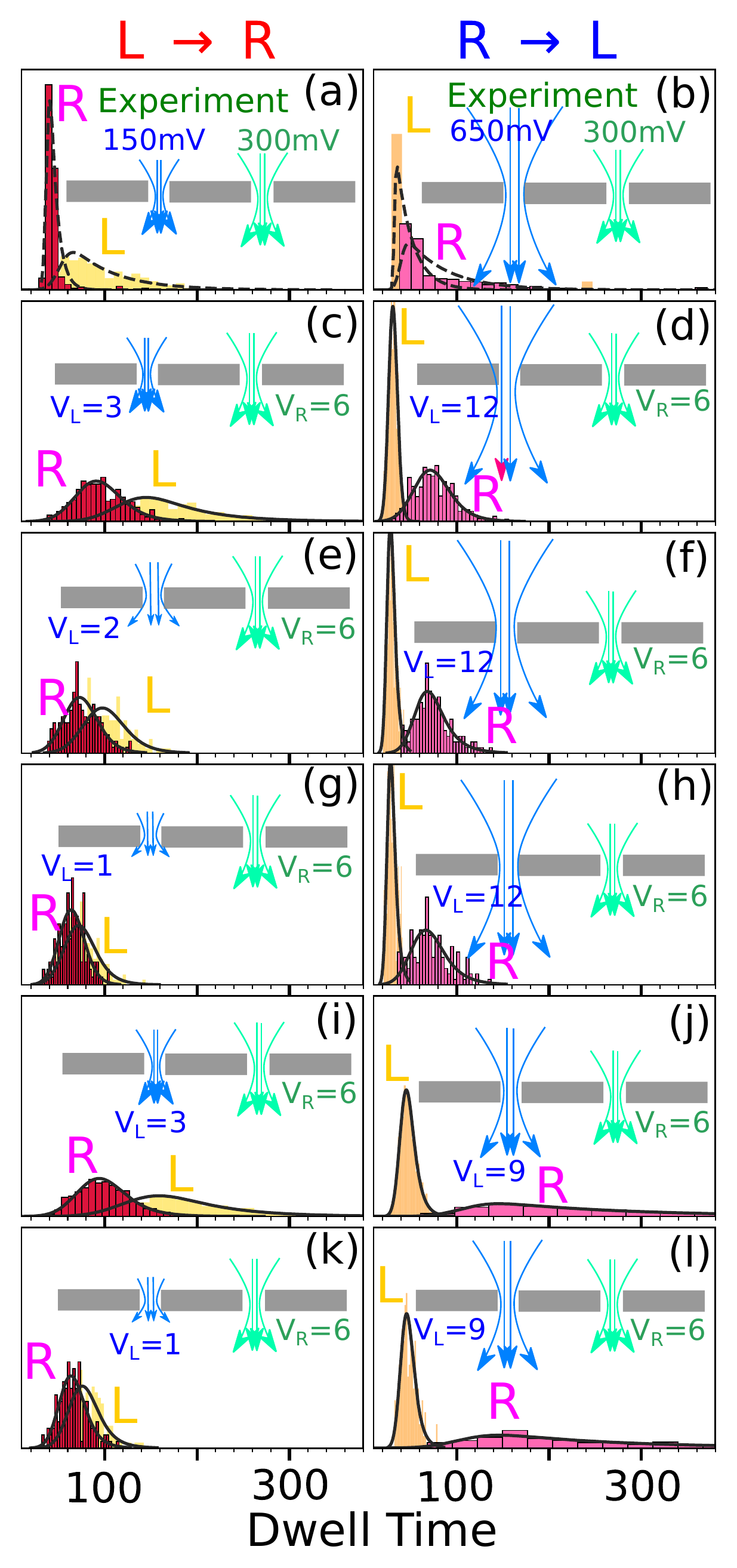}
\vskip -0.3truecm  
\caption{\small Cumulative dwell time distribution of the seven tags (sidechains) obtained from the experiment for (a) $L \rightarrow R$ and (b) $R \rightarrow L$ flossing (1st row). The rest of the rows (2nd - 6th) are simulation dwell time data for different combination of voltages $V_L$ and $V_R$ applied across the left and the right pore for the seven sidechains placed exactly at the same locations as that of the experiment. In each row the yellow/red (left column) and the orange/magenta (right column) dwell time histograms 
are obtained from the left/right pore in $L \rightarrow R$ and $R\rightarrow L$ directions.  
Schematics of the electrostatic force on the DNA in left/right pore are shown by the blue/green arrows (not to scale). The black envelops represent the exponentially modified Gaussian distribution fit of the dwell time histogram.} 
\label{Dwell}
\end{figure}
The electric field is calculated inside and in the vicinity of the nanopore exactly by solving Maxwell's equation with the proper boundary condition (supplementary section on E-field). The field is strongest inside the pore, extends but fades away quickly outside
the pores. By varying the voltage at the left pore to drive the DNA back and forth makes the process inherently asymmetric  
as translocating beads face different energy barriers for the ${L\rightarrow R}$ and the ${R\rightarrow L}$ translocations. This is reflected in the dwell time distribution for the charged
side-chains shown in Fig.~\ref{Dwell}. We have checked (not shown here) that the charged spherical tags also exhibit asymmetric dwell time distributions. 
During flossing the chain conformations are compressed leading to a relatively faster translocation process compared to the relaxation process that takes place at a much longer time scale\cite{Bhattacharya_Seth_2020,Seth-JCP-2020}. 
\par
We first discuss the general characteristics of the ${L\rightarrow R}$ translocation. In this case a charged tag translocates through the left pore against the field but the field favors the translocation through the right pore (Fig.~\ref{E-Field}). Thus, for the ${L\rightarrow R}$ translocation the dwell time should be broader at the left pore and sharper at the right pore. The widths of the distributions are reversed for the ${R\rightarrow L}$ translocation. However, the strength of the opposing field at the R-pore is stronger for the ${R\rightarrow L}$ and the favorable field at the L-pore is weaker which makes the distributions for the $L\rightarrow R$ and $L\rightarrow R$ different as reflected in Fig.~\ref{Dwell}.
\par
Each row in Fig.~\ref{Dwell} shows the dwell time distributions for the $L\rightarrow R$ and $R\rightarrow L$ respectively. The first row shows the experimental dwell time distribution (Fig.~\ref{Dwell}(a)-(b)) where $V_L=150$ mV and $V_R=300$ mV for $L \rightarrow R$ translocation so that $V_R/V_L=2.0$. Similarly, for the $R \rightarrow L$ translocation  $V_L=650$ and $V_R$ is kept the same and in this case $V_R/V_L=0.5$ (Please refer to the supplementary section on Experimental Method). As expected, the dwell times are consistent with the above discussion so that $W_L^{L \rightarrow R}$ is broader compared to $W_R^{L \rightarrow R}$. Likewise, for the ${R\rightarrow L}$ translocation this order gets reversed. However, one notices that  $W_R^{R \rightarrow L}$ is different than $W_L^{L \rightarrow R}$ as the biases are altered. During flossing the translocated chain at the left or right pore gets compressed to a different degree depending upon the strength of the downhill bias. Thus, when the voltage gets flipped, the 
degree of compression affects the unfolding and hence the speed of the translocation differently for ${L\rightarrow R}$ and ${R\rightarrow L}$ translocation. We have checked that a compressed configuration translocated faster than a fully equilibrated configuration~\cite{SS-Unpublished}. In the experimental protocol, there is an uncertainty in the voltage applied at the left pore. Thus, we have used several different combinations of the biases in the simulation studies to check how relative strength of the voltage at each pore affects the translocation process.
\par
 The next five rows of Fig.~\ref{Dwell}(c)-(d) - (k)-(l) are the dwell time distributions obtained from our simulation using charged side-chain tags. The second row (Fig.~\ref{Dwell}(c)-(d)) corresponds to the experimental parameters where we kept the ratio $V_R/V_L=2$ and $0.5$, the same for the $L \rightarrow R$
and $R \rightarrow L$ translocation as in the original experiment. Despite the simplicity of the model the simulation studies capture this asymmetry quite well excepting the simulation  $W_L^{L \rightarrow R}$ is slightly broader than the experimental one, nevertheless, the simulation predicts the qualitative trends.  We further observed that the asymmetry resembles closer to the experiment for the charged side chain tags than the charged spherical tags (not shown). It is worth noting that there are some unknown factors those are not
accounted for in the simulation, such as surface charges inside the pore, co-ion and counter ion movements, uncertainties in the applied voltage, and roughness of the pore. Therefore, we have further explored other combinations by first systematically reducing the bias at the left pore to $V_L^{L \rightarrow R} = 2$ (Fig.~\ref{Dwell}(e)-(f)), and $V_L^{L \rightarrow R} = 1$ (Fig.~\ref{Dwell}(g)-(h)). Reducing the bias at the left pore for the ${L \rightarrow R} $ translocation only changes the $W_L^{L \rightarrow R}$ without noticeably affecting the the distributions at the right pore. While changing $V_L$ for the $R \rightarrow L$ translocation causes the distributions drifting away from the experimental results. This qualitative agreement clearly shows that the charge of the side-chain tags is the dominating factor for the shape of the distributions. \par
\begin{figure}[ht!]
\includegraphics[width=0.49\textwidth]{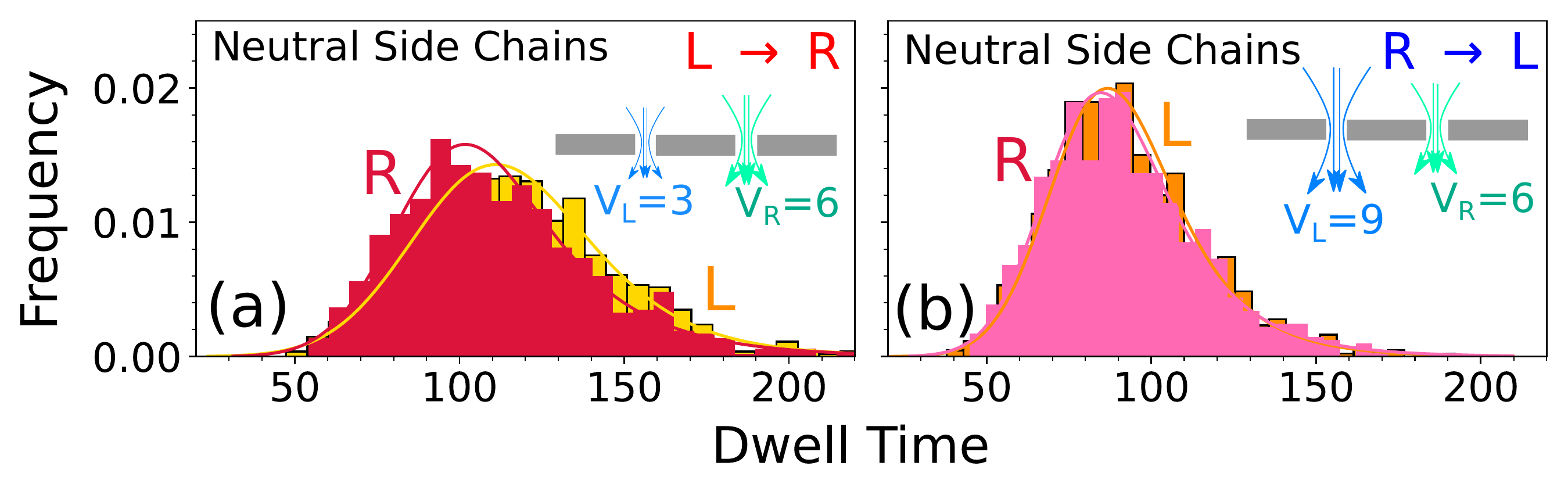}
\caption{\small (a) The cumulative dwell time distribution for the left and right nanopore for the neutral sidechain tags.  
Unlike the charged side-chain tags the asymmetry almost disappears.  }
\label{neutral}
\end{figure}
\begin{figure*}[ht!]
\includegraphics[width=0.9\textwidth]{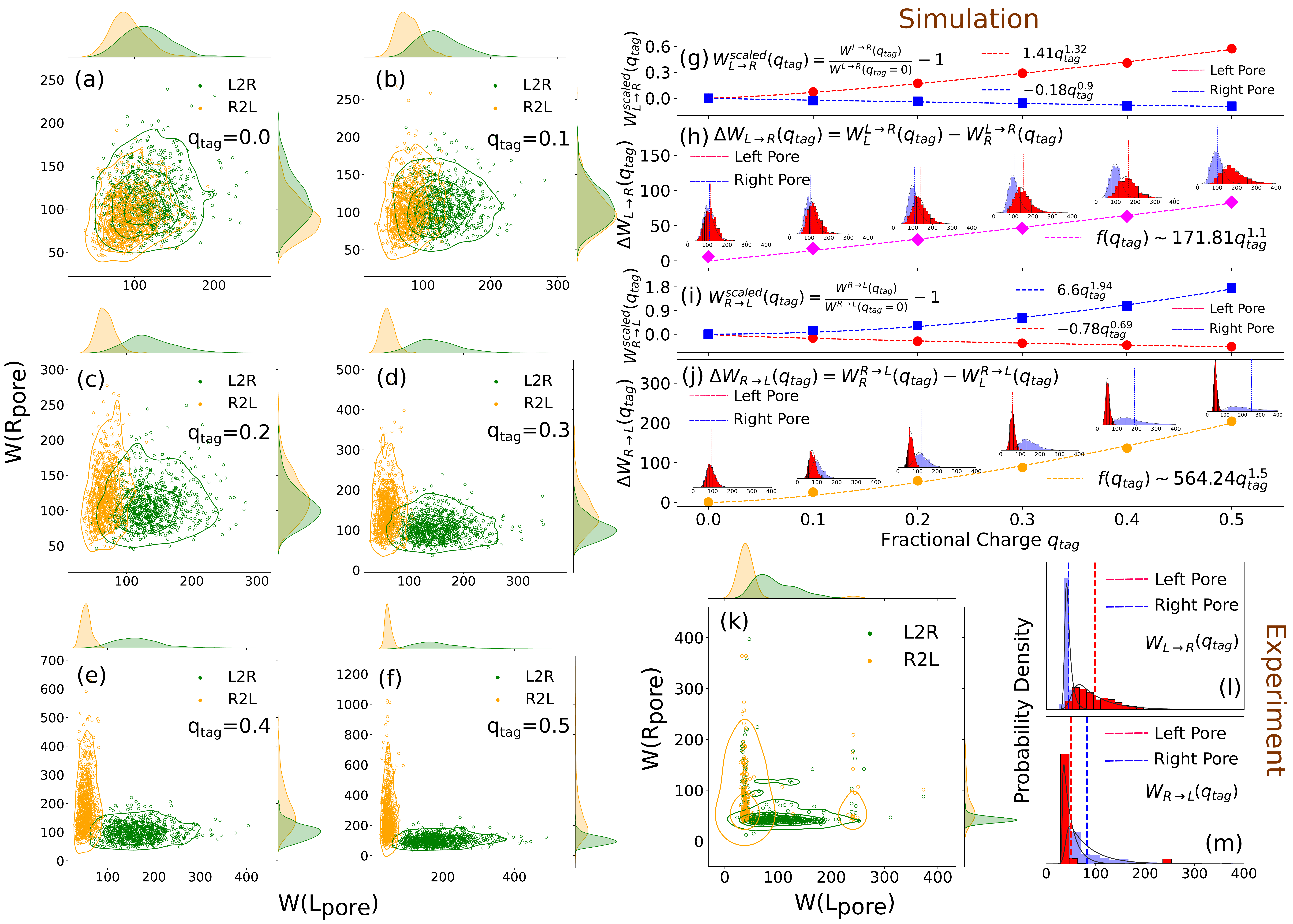}
\caption{\small Scatter plots of the cumulative dwell time for the neutral (a) and the partially charged sidechain tags (b)-(f) in the L/R pores for $L\rightarrow R/ R\rightarrow L$ scans (green/orange circles). Distributions of the corresponding dwell time are shown on the top and right hand axes. (g) The scaled cumulative dwell time asymmetry $\Delta W_{L\rightarrow R}$ at the left (blue squares) and right (red circles). The dotted lines (red and blue) are the corresponding nonlinear fits through the points which produce different exponents for the charge dependence. (h) The corresponding histograms of the points in (g) where the dotted vertical lines (red and blue) in each histogram represent the average dwell time. (i) and (j) are the same as in (g) and (h) excepting for the ${R \rightarrow L}$ translocation direction. The fitting exponents in (i) are also different than those in (h). (k) the experimental scatter plot to be compares with (a)-(f). (l) and (m) are the corresponding experimental  $\Delta W_{L\rightarrow R}$ and $\Delta W_{R \rightarrow L}$ (same as in Fig.~\ref{Dwell}(a) and (b) for comparison.}
\label{charge}
\end{figure*}
\begin{figure}[ht!]
\includegraphics[width=0.48\textwidth]{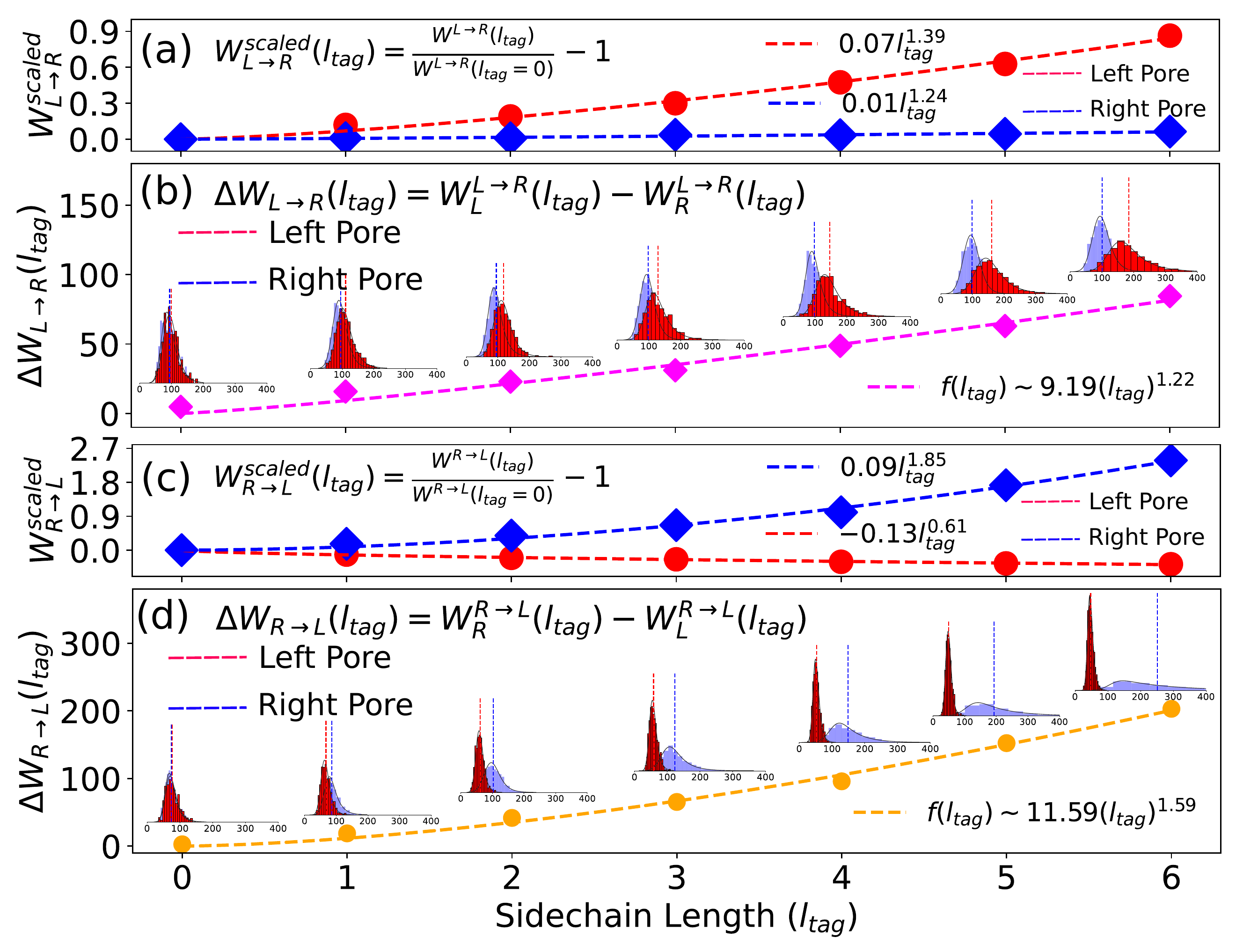}
\caption{(a) The scaled cumulative dwell time asymmetry $\Delta W^{scaled}_{L\rightarrow R}$ at the left (red circles) and right (blue squares) pore. The dotted lines are the nonlinear fits through the points which produce different exponents for the charge dependence. (b) The corresponding histograms of $\Delta W_{L\rightarrow R}(l_{tag})$ for different $l_{tag}$ locations as in (a) where the vertical red and blue dotted lines in each histogram are the average values of the dwell time. (c) and (d) are the same as in (a) and (b) for $\Delta W^{scaled}_{R\rightarrow L}$.
\label{mass}}
\end{figure}
We have checked that this asymmetry almost disappears for the neutral tags as shown in Fig.~\ref{neutral}. \par
{\bf How does the partial charge in side chains affect the dwell time ?} From the discussions of Figs.~\ref{Dwell} and \ref{neutral} it is clear that the 
charge of the tags is one of the key factors that controls the asymmetry of the dwell time for the  $L \rightarrow R$ and  $R
\rightarrow L$ translocation. 
The charge of lifts away the near degenerate $L \rightarrow R$ and $R \rightarrow L$ distributions for the neutral tags ($q_{tag}=0$) (Fig.~\ref{neutral}). Thus, we further explored the asymmetry of the dwell time distributions (Fig.~\ref{Dwell}) in terms of $\Delta W_{L \rightarrow R}(q_{tag})$ and  $\Delta W_{R \rightarrow L}(q_{tag})$ (Eqn.~\ref{dw}) and its normalized counterpart (Eqn.~\ref{sdw})
defined as 
\begin{subequations}
  \begin{equation}
      \Delta W_{L \rightarrow R}(q_{tag})  = \langle W_L^{L \rightarrow R}(q_{tag})\rangle - \langle W_R^{L \rightarrow R}(q_{tag})\rangle
  \label{dw} 
 \end{equation}
   \begin{equation}
  \Delta W^{scaled}_{L \rightarrow R}(q_{tag})= \frac{ \Delta W_{L \rightarrow R}(q_{tag})}{\Delta W_{L \rightarrow R}(q_{tag}=0)}-1.
  \label{sdw}
\end{equation}
\end{subequations}
for the charged sidechains shown in Fig.~\ref{charge}. Likewise, one can define  $\Delta W_{R \rightarrow L}$ by interchanging $L$ and $R$.
The differential functions $\Delta W^{scaled}_{L \rightarrow R}(q_{tag})$ and  $\Delta W^{scaled}_{R \rightarrow L}(q_{tag}) \rightarrow 0$ as
the charge of the sidechain protein tag $q_{tag} \rightarrow 0$ and brings out more effectively the local effects of each pore on $q_{tag}$.
The scattered plots of Figs.~\ref{charge}(a)-(f) clearly brings out how the isotropy ($q_{tag}=0$) is broken and continuously evolve to acquire characteristics of the local effects from the left and the right pore as the charge of each sidechain is increased. This is an excellent demonstration of the importance of simulation studies to understand the corresponding experimental data shown in Fig.~\ref{charge}(k). By comparing Fig.~\ref{charge}(k) with the set in Figs.~\ref{charge}(a)-(f) one can infer that not only the sidechain protein tags are charged, one can also estimate the partial charge content of the tags. \par
An important aspect of the experimental set up (Fig.~\ref{E-Field}) is that variation of the electric field occurs at the L-pore while the voltage at the R-pore is kept constant which results in the asymmetries in dwell time. In Fig.~\ref{charge}(g)-(j) we further explore in detail the variations of dwell time at the L/R pore as a function of the charge of the sidechain motifs those  can be easily understood by looking at the field directions (Fig~\ref{E-Field}) at the L/R for both $L/R \rightarrow R/L$ translocation. 
For example in Fig.~\ref{charge}(g) the E-field at the L/R pores are antiparallel/parallel for the $L \rightarrow R$ translocation that explains why 
${W}^{scaled}_{L \rightarrow R}$ increases/decreases at the L/R-pore. Fig.~\ref{charge}(g) then immediately explains the monotonic increase of 
$\Delta W_{R \rightarrow L}(q_{tag})$. Likewise ${W}^{scaled}_{R \rightarrow L}$ increases/decreases at the R/L-pore (Fig.~\ref{charge}(i)) and explains the monotonic increase of the differential function $\Delta W_{R \rightarrow L}(q_{tag})$ (Fig.~\ref{charge}(j)). 
Furthermore, we find that the data in Fig.~\ref{charge}(h) and (j) can be fitted with a power law dependence 
$\Delta W_{L \rightarrow R}(q_{tag}) \sim A_{LR}q^{\alpha_{LR}}$ (and likewise, for $\Delta W_{R \rightarrow L}(q_{tag})$), where both the prefactors $A_{LR}$ and  $A_{RL}$, and the exponents $\alpha_{LR}$ and $\alpha_{RL}$ at the L/R pore are in general different and nonuniversal, and depends on the details of the parameters (please refer to the annotations in Fig.~\ref{charge}). We strongly this power law dependence can be potentially used to determine effective charge of a motif in an experimental scan and can potentially discriminate tags by their charge contents. \par
{\bf How does the length of the side chains affect the dwell time ?}
In general protein tags of different lengths can be present in a long DNA-strand. Then a dual nanopore device will be able to detect the presence of different protien tags. In the previous section we studied the effect of the magnitude of the partial charge keeping the length of the sidechain protein tags the same. We have carried out similar analysis replacing charge $q$ by the length of the tag $l_{tag}$ in Eqn.~\ref{dw}
and Eqn.~\ref{sdw} keeping the charge per bead the same as shown in Fig.~\ref{mass}. We observe power law behavior for both $W^{scaled}_{L \rightarrow R}(l_{tag})$ and
$W^{scaled}_{R \rightarrow L}(l_{tag})$ (Fig.~\ref{mass}(a),(c)) and monotonic increase for $\Delta W_{L \rightarrow R}(l_{tag})$ and $\Delta W_{R \rightarrow L}(l_{tag})$ (Fig.~\ref{mass}(a),(c)).

It is worthwhile to observe that significant deviations form linearity as a function of $q_{tag}$ and $l_{tag}$  are observed. This is expected due to a combined effect of tension propagation  and varied degree of inertial and field effects at each pore due to chain connectivity. 
\par
\begin{figure}[ht!]
  \includegraphics[width=0.47\textwidth]{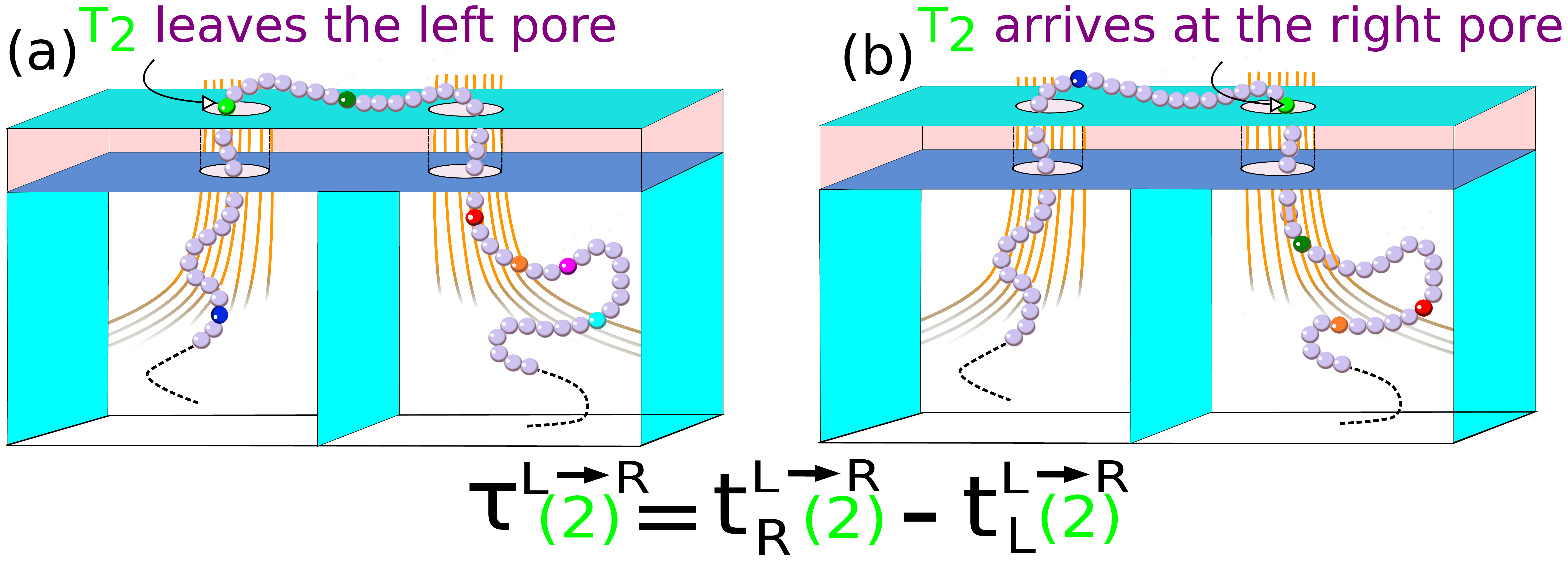}
  \vskip -0.25truecm  
\caption{\small \label{tof} Illustration depicts the TOF of ${\bf\color{green} T_2}$ is measured as the time taken to reach to left pore from right pore for $R \rightarrow L$ motion.}
\label{tof-fig}
\end{figure}
\begin{figure}[ht!]
\includegraphics[width=0.47\textwidth]{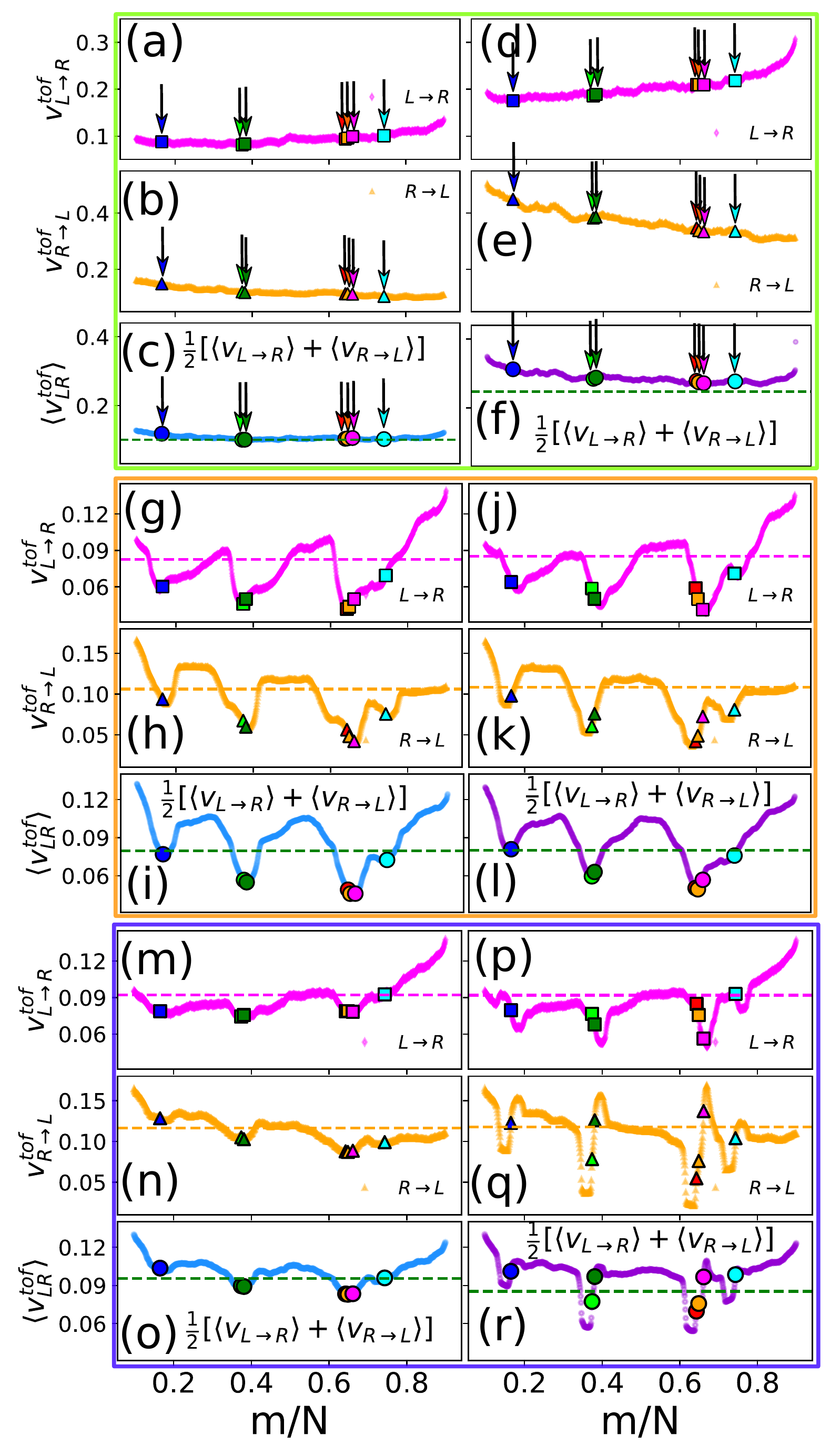}
\vskip -0.3 truecm
\caption{ TOF velocity for neutral ((a)-(c)) and charged tags ((d) - (f)) with the same mass as that of the chain monomers. The locations of the tags are indicated by arrows. The positive slopes in $L \rightarrow R$ ((a), (d)) and negative slope in $R \rightarrow L$ ((b),(e)) denote the direction of the TP along the chain; (g) - (i) and (j) - (l) represent the same but for heavier (6m) neutral and charged (0.5$q$) {\em spherical} tags. The heavier tags introduces nonuniformity in the TOF velocity; (m) - (o) and (p) - (r) represent the same as that of (g) - (i) and (j) - (l) respectively but for the neutral and charged {\em side-chain} tags. The side-chains exhibit more discriminating features compared to the spherical tags of the same mass.
\label{velocity}}
\end{figure}

{\bf Velocity of the tags from the time of flight (TOF) data:}~Compared to a single
nanopore, in a dual nanopore setup the velocity of the tags are calculated more accurately from the TOF measurements \cite{SS1}
defined as the time taken by a monomer/tag of index $m$ as it leaves one pore and reaches the other pore during its voyage across the pore 
separation $d_{LR}$ defined as
\begin{subequations}
 \label{tof-eqn}
\begin{gather}
\tau^{L \rightarrow R}(m) = t_{R}^{L \rightarrow R}(m) - t_{L}^{L \rightarrow R}(m) \\
\tau^{R \rightarrow L}(m) = t_{L}^{R \rightarrow L}(m) - t_{R}^{R\rightarrow L}(m) 
\end{gather}
\end{subequations}
Here $t_{R}^{L \rightarrow R}(m)$  and $ t_{L}^{L \rightarrow R}(m)$ are the arrival and exit time at the right and left pore for the ${L \rightarrow R}$ translocation. By flipping $L$ and $R$ we get $\tau^{R \rightarrow L}(m)$. This is demonstrated in Fig.~\ref{tof-fig}. 
The TOF flight velocity for a monomer/tag with index $m$ then can be easily  obtained from measurement of 
$\tau^{L \rightarrow R}$ and $\tau^{R \rightarrow L}$ using the known distance $d_{LR}$.
\begin{subequations}
  \label{tof_v-eqn}
\begin{gather}
v^{tof}_{L \rightarrow R}(m)=d_{LR}/\tau^{L \rightarrow R}(m) \\
v^{tof}_{R \rightarrow L}(m)=d_{RL}/\tau^{L \rightarrow R}(m) 
\end{gather}
\end{subequations}
The TOF flight measurements can be experimentally obtained from the current blockade data. However, because the tags are in general of different mass, charge, and volume they introduce nonuniformity in the velocity along different portions of the chain which is difficult to access experimentally. The simulation studies show that charged tags experience different forces compared to the dsDNA chain. This results in a nonuniform velocity profile along the chain which can be qualitatively understood using nonequilibrium tension propagation theory due to Sakaue~\cite{Sakaue_PRE_2007} and recently demonstrated in a single and dual nanopore setup~\cite{SS1,SS2}. In the following section we analyze how the mass and charge of the tags affect the velocity profile of the entire chain. One of the major goal of these simulation studies is to develop a fundamental understanding of the 
piecewise translocation process. Thus here in addition to mimicking the parameters used in the experiment, we study several other 
variations of the tag characteristics to decipher the effects of the mass and charge of the tags which create the nonuniform velocity profile along the chain shown in 
Fig.~\ref{velocity}. We learn the following from a closer look at Fig.~\ref{velocity}. \par
$\bullet$~As a reference first we show the results for the neutral and charged tags of the same mass as that of the dNA monomers (Fig.~\ref{velocity}(a)-(f)). Here, we observe that velocity increases/decreases as a function of the monomer index $m$ for
$L \rightarrow R$ and  $R \rightarrow L$ translocation similar to a homopolymer~\cite{SS2}. Inclusion of the charge at protien tag locations do change the overall profile as expected but do not introduce significant nonuniformity in the velocity profile (Fig.~\ref{velocity}(d)-(f)). 
\par
$\bullet$~It is only when tags are more massive compared to the monomer beads we observe a huge nonuniformity in the
velocity profile of the entire chain with local minima roughly at the location of the indices of spherical tags (Fig.~\ref{velocity}(g)-(i) and (j)-(l)). 
Thus we conclude that it is the inertia of the tags responsible for the local minima. A comparison of Fig.~\ref{velocity}(g)-(i) and (j)-(l) shows that replacing 
neutral spherical tags by charged spherical tags does not alter
the profile confirming the role of inertia for the case when tags are massive but have the same volume as that of the monomer
beads. \par
$\bullet$~Replacing neutral spherical tags
(Fig.~\ref{velocity}(g)-(i)) by neutral side-chains of the same total mass reduce the nonuniformity of the velocity profile. This shows the interaction of the extended tags with the electric field beyond but in the vicinity of the pore and the entropy can make a significant difference. \par
$\bullet$~From an inspection of  Fig.~\ref{velocity}(g)-(i) and (j)-(l) we further observe that the TOF velocity has the
capability of discriminating density distribution of the tags. There
are two isolated tags at each end and two groups of tags - a group of two ($T_2,T_3$) and a group of three ($T_4,T_5,T_6$) in the system that we studied. For the {\em neutral} spherical tags for the $L \rightarrow R$ translocation the group of tags on which the tension front hits last lies in the minimum of the velocity profile ($T_4$ and $T_2$ in Fig~\ref{velocity}(g)) while for the $R \rightarrow L$ translocation
the location of the minima gets reversed ($T_6$ and $T_3$ in Fig~\ref{velocity}(h)). This sequence of $v_{\rm TOF}(m) \sim m$ is reversed when {\em spherical neutral} tags become charged as we compare  
Figs.~\ref{velocity}(g) and (j). 
\begin{figure}[ht!]
\includegraphics[width=0.48\textwidth]{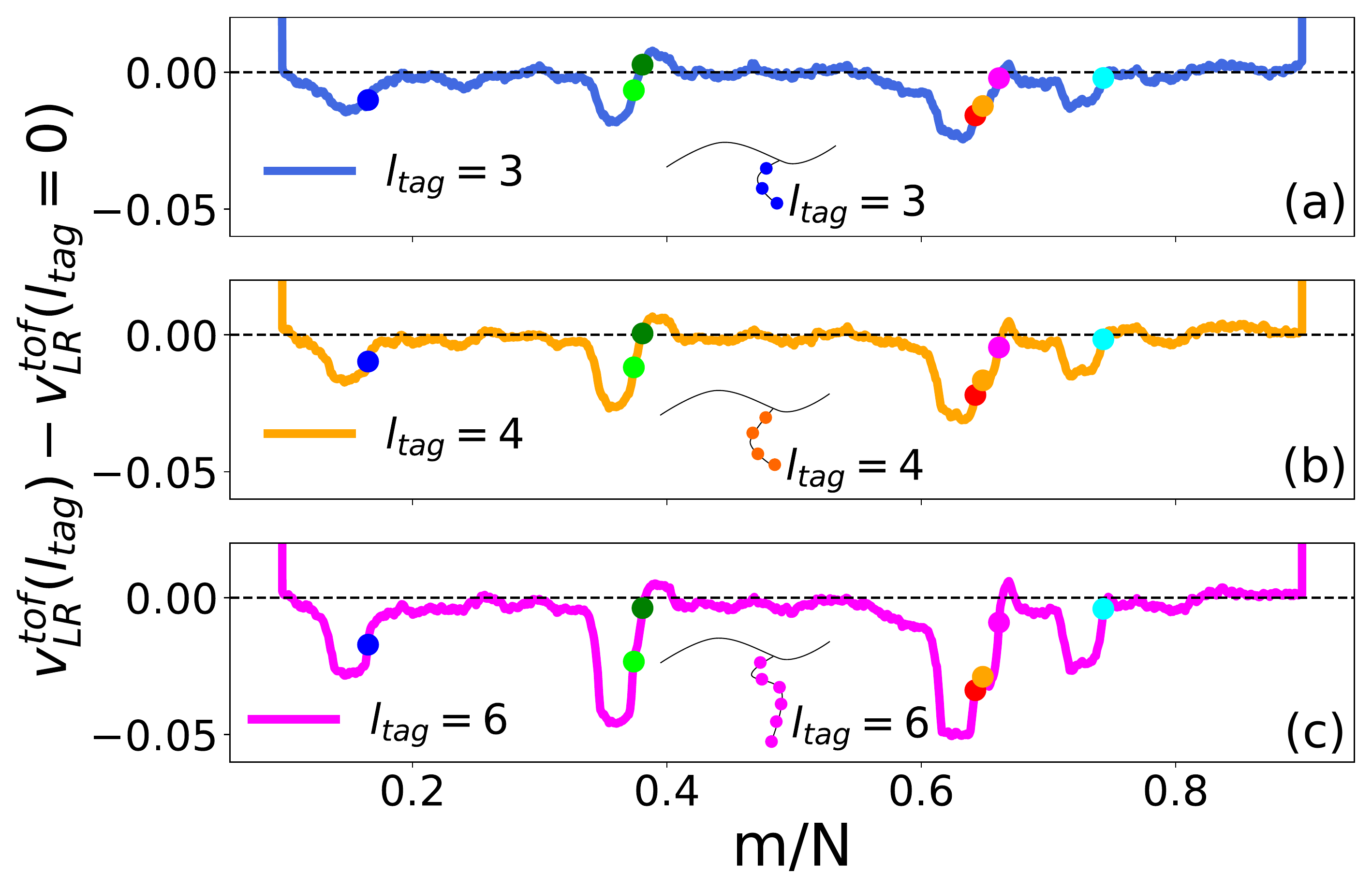}
\vskip -0.3 truecm 
\caption{Variation of velocity profile of the entire chain as a function of sidechain tags of different length $l_{tag}$. To extract the dependence of the sidechain protein tags we have subtracted the velocity of the velocity of the homopolymer ($l_{tag}=0$). (a), (b) and (c) correspond to $l_{tag}=3,4$ and $6$ respectively. The increasing local variations around position of each sidechain tag is evident. It is also worth observing that the last tag in all three cases has the same velocity of the entire chain and therefore, can be used as a reference velocity for the entire chain.
\label{inertia}}
\end{figure}\par
$\bullet$~Finally, we compare the velocity profiles for charged spherical tags (Fig.~\ref{velocity}(j)-(l)) and charged side-chain tags (Fig.~\ref{velocity}(p)-(r)) which are relevant to understand the corresponding experimental data. Evidently, the velocity profile of the charged side-chains is resolved better compared to the spherical tags with detailed separation distances. This is due to the interaction of the side-chains which are more extended than the spherical tags hence interacting with a larger region containing the electric field. In this case not only the segment of the side-chain tag inside the pore sense the electric field inside the pore, but the
segment lying outside in the immediate vicinity of the pore feels the electric field which extends from inside to the vicinity of the pore. This {\rm pre-sensing} of the nanopore by the side-chains provides the velocity a more intricate structure as seen in Fig.~\ref{velocity}(p)-(r). It is also worth observing that the discrimination is more prominent for the $R \rightarrow L$
translocation which then can be used to our advantage to decipher the characteristics of the tags.\par
{\bf Sidechain length dependence of the velocity profile:~}We close the discussion of the velocity variations by showing its dependence on the length of the sidechains $l_{tag}$. Previously in Fig.~\ref{mass} we showed the variation of the dwell time on 
$l_{tag}$. Since the velocity is determined experimentally from the TOF flight data, it is worthwhile to study how the sidechains alter the velocity profile. Fig.~\ref{inertia} shows systematic increase  of the inertial effect as the tag length (and hence mass) is increased.
To extract the variation we have subtracted out the velocity of the homopolymer chain without the protien sidechain tags.
An important observation in Fig.~\ref{inertia}(a)-(c) is that the very last tag has the velocity of the entire chain. This information can be useful to extract the velocity data experimentally using the last tag as the reference.
We also note that inertial effect become more prominent with increasing length of the sidechains $l_{tag}$. 
Thus we believe that the degree of velocity variation can be used to resolve protein tags of different mass and length (possibly of different origin) present along the dsDNA construct.
\par
{\bf Simulation versus experimental time scale:} In the Brownian Dynamics (BD) simulation we use a chain of $N=1024$ beads ($L=1024\sigma$) that corresponds to a 48500
bp long dsDNA construct. This leads to the diameter of each bead $\sigma = 48500/1024 \approx 47 $ bp $\approx 47*0.34 = 16$ nm. We 
now calculate the unit of BD time scale to relate the velocity of the chain under bias to compare with those from experiments. 
\par
{\bf Mass of a base pair}:~The average mass of a DNA base-pair  $m_{DNA} \simeq 650 \;\rm{amu} = 1.67\times 10^{-27} {\rm kg}\;\times 650 \simeq 1.1 \times 10^{-24} \;
  kg$. This estimate can be obtained by looking at the chemical structures of the $A-T$ and $G-C$ bonds, and accounting for the molecular weights of deoxy-Ribose and the phosphate group both contributing to the dsDNA mass. This will allow us to get the time unit for the BD simulation from the following equation.
  \begin{subequations}
  \begin{equation}
  \tilde{t}_{BD}=  \sqrt{\frac{m_{DNA}\sigma^2}{\epsilon} } \simeq
  0.26 \;{\rm ns}
  \label{t_BD}
 \end{equation}
\begin{equation}
    \therefore \tilde{v}_{BD} = \frac{\sigma}{\tilde{t}_{BD}} =
    \sqrt{\frac{k_BT}{m_{DNA}}} \simeq 61.0 \;{\rm m/s}.
    \label{v_BD}
  \end{equation}
\end{subequations}
Here, we have used $\epsilon \simeq k_BT \simeq 4200 \times 10^{-24}$ J per monomer bead. Thus, $\tilde{v}_{BD}$ is independent of $\sigma$ and only depends on the interaction strength and the mass of the beads. From simulation, typical dimensionless values of $\tilde{v}_{TOF}\approx 0.1$ (from Fig.~\ref{velocity}) which translates to the actual velocity =
$\tilde{v}_{TOF} \times 60 \;{\rm m/s} \approx 6.0\;{\rm m/s} = 600\;{\rm mm/s}$. It is well known that the BD simulation with implicit solvent makes the time
scale faster depending upon the degree of coarse graining. Considering
we have translated 48500 bp to 1024 coarse-grained beads (a factor of
$\approx$ 50), this simulation in
actual solvent would translate to the velocity $ \tilde{v}_{TOF}/50 \approx$ 10 mm/s, which is the typical order of magnitude of velocity for the dual nanopore experiments~\cite{Small2,Small3}.
This is the reason why the BD simulation is capable of reproducing the same experimental trend qualitatively. It is expected that if one would carry out a more expensive calculation with explicit solvents and with a longer chain the agreement will be similar. Roughly speaking the BD simulation captures the physical phenomena albeit at a faster time scale depending upon the degree of coarse-graining. 
\par
{\bf Translating simulation bias to the experimental bias \& P\'{e}clet number:}~
The average velocity of the chain will depend on the applied bias. In
the original experiments~\cite{Small1,Small2,Small3} the bias at the
left pore is varied from $F\sigma \approx k_BT - 10 k_BT$, while the
bias at the right pore is larger. The biases used in the BD simulation
should be commensurate with this. Please note that in Fig.~\ref{E-Field} we used the same factor of 50 in translating 150 mV
to 3 BD simulation units. We will now show the internal consistency by calculating the P\'{e}clet number as outlined below.\par
When the voltage bias is increased from $k_BT \simeq 1$ to 10, the diffusive motion changes over to the drift. Thus  several authors used 
P\'{e}clet number
\begin{equation}
  P_e=\frac{\tau_{\rm relax}}{\tau_{\rm trans}}.
\end{equation}
to compare the
applied bias used in the simulation~\cite{Saito_2012,deHaan,Stein}
with those in the experiment.
Here $\tau_{\rm relax}$ and $\tau_{\rm trans}$ are the
relaxation and translocation time for the translocating polymer, thus
is a measure of diffusive versus drift motion. A comparison of P\'{e}clet number can provide useful information in this context. Previous
studies~\cite{Saito_2012,deHaan,Stein,Storm} were done in reference to a
single nanopore. We use a similar argument for the dual nanopore
system. However, it is worth pointing out that during flossing the
chain does not escape completely, rather a major segment of the chain (90\% in our simulation) is scanned back and forth. Therefore, initial conformations of the chain those translocate through the dual
nanopore are far from equilibrium and are different from well established studies carried out in the context of a single nanopore where a scaling exponent is sought for the driven
translocation for an equilibrated initial chain. The compressed configuration in our study unfolds and translocates faster depending on the degree of compression. Flossing a $\lambda$-phage DNA-construct with seven tags in a dual nanopore setup is performed under bias
forces $\Delta F_{LR}$ ranging from $150-650$ mV~\cite{Small2,Small3} which results in a typical TOF velocity $v_{TOF}^{expt} \simeq 0.77$ mm/s (Supplementary Material Table
S3) \cite{Small3}. Hence, for a $16.6\; {\rm \mu m}$ $\lambda$-phage dsDNA
$\tau_{trans}^{expt} \simeq (16600/0.77)\;{\rm \mu s}\approx
0.02\;s$. \par
For $L=16.6\; {\rm \mu m}$ $\lambda$-phage dsDNA used in the dual nanopore experiment~\cite{Small2} we use the formula
by Smith {et al.}~\cite{Smith} $D_{bulk}^{expt} = 2.38/L^{0.608} =
0.43\;{\rm \mu m^2/s}$.
and the formula for the bulk radius of gyration $(R_g)_{bulk}^{expt} = 0.146 L^{3/5}$. This gives
$(R_g^2)_{bulk}^{expt}\simeq 0.621 \;{\rm \mu m^2}$ that gives
$\tau_{relax}=(0.621/0.5) s =1.24 s$.
Thus, for the double nanopore experiment  P\'{e}clet number is 
$P_{Exp}^{DNP} = 1.24s/0.02s \approx 60$.
\par
Now we get the P\'{e}clet number for the BD simulation using $\sigma =16\; {\rm nm}$ and $\tilde{v}_{TOF} \simeq 0.1$.
The average translocation time from multiple scans
$\tau_{trans}^{sim}=1024/0.1 \simeq 10240$.
To get the BD simulation relaxation time we use the relation $\tau_{relax} \sim B^2 \gamma N^{2.2} = 0.153\times (1024)^{2.2} = 643057$ in $k_B T/\sigma = 1$
unit~\cite{deHaan} from where we obtain $P_{Sim}^{DNP}=643057/10240 \simeq 63 \approx P_{Exp}^{DNP} \approx 60$.
Thus, this agreement of the P\'{e}clet numbers from experiment and simulation further justifies and closes the loop why the BD simulation studies capture the essential features of flossing in a dual nanopore device and give confidence to use this model for analysis of a more complicated mixed system of tags {\em in silico}.

\par
{\bf Concluding remarks:}~We have developed novel BD simulation strategies whose overarching goal is to extract the the underlying physics of the dual nanopore translocation at sub-nanometer length scales hard to obtain experimentally and hence 
improve the accuracy of the locations of protein tags on dsDNA constructs based on the details as revealed from the results obtained from the CG model. The simulation strategies are also capable of predicting possible variations of the device characteristics of the dual nanopore system to improve its accuracy. In the BD simulation we varied the magnitudes of local electric fields at each nanopore and demonstrated  that both the average dwell time and degree of asymmetries due to opposing and favoring local fields follow power laws as a function of the charge as well as the length of the protein tags, albeit with different exponents and amplitudes. 
Establishment of such a result will be useful to analyze experimental data as one can study how the shapes of the dwell time distribution get altered under different electric fields as well as the characteristics of the protein tags. 
The time evolution  of a flossed dsDNA subject to repeated scans needs to be understood in terms of nonequilibrium statistical mechanics. We have explained a variety of scenarios in terms of polymer physics concepts and nonequilibrium tension propagation theory, such as, how the fine structures of the velocity profile of the entire chain are altered due to the presence of the protein tags. 
Finally, a direct relevance of the model to experimental results is the observation that the coarse-graining length factor which is the ratio of the actual length to the simulation chain length provides a guide how to compare  the experimental velocity to the simulation velocity. This argument is validated by comparing the P\'{e}clet numbers. Our studies demonstrate that protein tags of different biological origins can be discriminated in terms of their physical characteristics enabling the simulation protocols to have huge potential application in genomics.
\par
{\bf Acknowledgment:}~The research has been supported by the grant
number 1R21HG011236-01 from the National Human Genome Research
Institute at the National Institute of Health. All computations were carried out using STOKES High Performance Computing Cluster at UCF. The simulation movie is generated using the Visual Molecular Dynamics package~\cite{VMD}.\par
{\bf Author contributions:}~A.B. directed the project. A.B. and S.S planned the simulation studies and wrote the first draft. S.S performed all the simulation studies and prepared all the graphs. A.R. and W.R. provided the experimental data. B. D., R.S., and W.R. contributed to the final draft of the manuscript.

\vfill
\end{document}


\title{Supplementary Material for  ``Discriminating protein tags on a dsDNA construct using a Dual Nanopore Device''}   
\author{Swarnadeep Seth$^1$}
\author{Arthur Rand$^4$}
\author{Walter Reisner$^2$}
\author{William B. Dunbar$^4$}
\author{Robert Sladek$^3$}
\author{Aniket Bhattacharya$^1$}

\altaffiliation[]
{}
\affiliation{$^1$Department of Physics, University of Central Florida, Orlando, Florida 32816-2385, USA}
\affiliation{$^2$Department of Physics, McGill University, 3600 rue university, Montreal, Quebec H3A 2T8, Canada}
\affiliation{$^3$Departments of Medicine \& Human Genetics, McGill University, Montreal, H3A 0G1, Canada}
\affiliation{$^4$Nooma Bio, 250 Natural Bridge Dr, Santa Cruz, CA 95060, USA}

\maketitle

\noindent
Experimental results for tag dwell-times in our dual-pore device are obtained using previously published methods.   Device fabrication is described in Zhang \emph{et al.}\,\cite{nooma1}, use of FPGA protocol for efficiently forming tug-of-war states is described in Li \textit{et al.} 2019\,\cite{nooma2}, basic multi-scanning protocols and tag blockade analysis are described in Li \emph{et al.} 2020\,\cite{nooma3} and zoom multi-scanning and recapture protocols are described in Rand \emph{et al.} (\cite{nooma4}, note this manuscript is currently under review and available on BioRxiv).  
\\
\\
Briefly, the labeled DNA constructs used are based on $\lambda$-DNA with 90\,nt long oligo-flap labels inserted at the recognition sites of the nicking enzyme Nt.BbvC1.  The labels are generated by using nick-translation to insert azide bearing nucleotides at the Nt.BbvC1 recognition sites; these are then coupled to 90\,nt long DBCO bearing moieties via copper-free click chemistry \cite{nooma4}.  The labeled constructs are introduced into a dual-pore device in 2\,M LiCl buffer containing two 20-30\,nm diameter pores fabricated via focused ion-beam (FIB) \cite{nooma1}.  The pores have a spacing of around 500\,nm.   Each pore is coupled to a separate microchannel that can be independently biased with respect to a common reservoir giving access to both pores.  The current through each pore is measured using a dual channel recorder (Molecular Device Multi-clam 700B) interfaced to a Field-Programmable Gate Array (FPGA, PCIe-7851R).  The FPGA can apply active logic to exploit feedback signals, in the form of measured currents at each pore, to dynamically adjust the dual pore biasing during a translocation event.  The signals are sampled at 250\,kHz and filtered at 10\,kHz.  Specifically, the results obtained correspond to the following parameters:  pore 1 has a diameter of 23.9\,nm, pore 2 has a diameter of 20.4\,nm and the pore-to-pore spacing is 561\,nm (obtained from SEM).
\\
\\
The labeled $\lambda$-DNA is captured in a tug-of-war state \cite{nooma2} and then subjected to a multi-scanning protocol where the molecule is driven back and forth between the pores (\cite{nooma3,nooma4}, which we term ``flossing").  In a dual-pore device, when a molecule trapped in a tug-of-war state, the translocation speed and polarity is determined fundamentally by the magnitude and sign of the biasing differential between the pores.  In our scanning protocol, to achieve either L-R or R-L directed motion, the voltage at one pore, which we will term pore 2, is held fixed (at 300\,mV) and the voltage at pore 1 is varied between a low (150\,mV, for L-R motion) and high level (650\,mV, for R-L motion) with respect to the pore 1 voltage.  In order to vary the scan polarity in a systematic way to achieve multi-scanning, the FPGA is then programmed to count the number of tags that pass through a pore (say pore 2 for a translocation in the L-R direction), and then reverse the translocation polarity from the L-R direction to the R-L direction when a fixed number of tags have passed through the pore.  The number of tags that pass through pore 1 is then counted and the translocation polarity flipped when a fixed number have passed through.  Repetition of these counting cycles at pore 1 and 2 leads to repetitive back-and-forth cycling of the chain.  In the zoom multi-scanning protocol \cite{nooma4}, the number of tags that are allowed to pass through pore 1 and 2 is gradually increased, leading to scanning of successively larger portions of the chain.  In addition, the FPGA can recapture the same chain to perform multiple multi-scanning cycles on the same chain \cite{nooma4}.  The tag results shown correspond to 18 separate $\lambda$-DNA molecules captured and 46 recapture events.   The translocation velocity $v$ can be measured by finding the time-of-flight (TOF) for a tag to move between the pores: $v=\langle TOF \rangle /d$ where $d$ is the pore-to-pore spacing and $\langle TOF \rangle$ is the average TOF over all tags measured for a given captured molecule (including all individual recaptures for the given molecule).  Averaging the translocation velocity further over all molecules captured gives:  $v_{L-R}=0.99\pm0.01$ (for L-R direction) and $v_{R-L}=2.23\pm0.05$ (for R-L direction).  Error given corresponds to error on the mean over all 18 molecules captured.
\\
\\
The tag-blockades are analyzed on individual scans using the approach developed in \cite{nooma3} coded in Matlab (see in particular supplementary methods for this article).   First, the positions of the tags along the scan are identified using a peak-finding algorithm.   Next, we model the varying current background arising from capacitance transients introduced by flipping the voltage.  To do this, a fixed region of around 400\,$\mu$s is removed about the position of each identified tag and then the remaining signal, giving purely the varying current background, is fit to a spline.  The spline model for the varying background is then substracted from the signals for pore 1 and pore 2.  The individual blockade profiles for each tag are then modeled via the fitting function:
\begin{equation}
I(t)=I_{b1}+I_{b2} t-\frac{I_o}{2} \left[\mbox{erf}\left(\frac{t-(t_o-\Delta t/2)}{\sqrt{2} \sigma}\right)-\mbox{erf}\left(\frac{t-(t_o+\Delta t/2)}{\sqrt{2} \sigma}\right) \right]
\end{equation}
This form is based on convolving a box of width $\Delta t$, height $I_o$ centered at position $t_o$ with a Gaussian of width $\sigma$.  In the limit that $\Delta t \gg \sigma$, the model yields a box-like profile.  In the limit that $\Delta t \ll \sigma$, the model yields a Gaussian-like profile.  The profile width at half-maximum is reported as the pore dwell-time (note that the width at half-maximum is obtained numerically as a function of $\Delta t$ and $\sigma$ from the blockade model).   The parameters $I_{b1}$, $I_{b2}$, which specify a constant and linear term, help characterize residual varying background at the tag-position.  The 46 recapture events obtained from the 18 molecules captured corresponds with multi-scanning to 2256 measured dwell-time pairs (e.g. for pore 1 and pore 2).